\documentclass[modern]{aastex631}

\usepackage{amssymb, stmaryrd}	
\usepackage{upgreek}
\usepackage{xcolor}

\begin{document}

\title{Visual Orbits of Wolf-Rayet Stars I: The Orbit of the dust-producing Wolf-Rayet binary WR\,137 measured with the CHARA Array}

\author[0000-0002-2806-9339]{Noel D. Richardson}
\affiliation{Department of Physics and Astronomy, Embry-Riddle Aeronautical University, 3700 Willow Creek Rd., Prescott, AZ 86301, USA}

\author[0000-0001-5415-9189]{Gail H. Schaefer}
\affiliation{The CHARA Array of Georgia State University, Mount Wilson Observatory, Mount Wilson, CA 91203, USA}

\author[0000-0002-1722-6343]{Jan J. Eldridge}
\affiliation{Department of Physics, University of Auckland, Private Bag 92019, Auckland 1142, New Zealand}

\author{Rebecca Spejcher}
\affiliation{Department of Physics and Astronomy, Embry-Riddle Aeronautical University, 3700 Willow Creek Rd., Prescott, AZ 86301, USA}

\author{Amanda Holdsworth}
\affiliation{Department of Physics and Astronomy, Embry-Riddle Aeronautical University, 3700 Willow Creek Rd., Prescott, AZ 86301, USA}

\author[0000-0003-0778-0321]{Ryan M. Lau}
\affiliation{NSF's NOIRLab, 950 N. Cherry Avenue, Tucson, Arizona 85719, USA}


\author[0000-0002-3380-3307]{John D. Monnier}
\affiliation{Astronomy Department, University of Michigan, Ann Arbor, MI 48109, USA}
\author[0000-0002-4333-9755]{Anthony F. J. Moffat}
\affiliation{D\'epartement de physique, Universit\'e de Montr\'eal, Complexe des Sciences, 1375 Avenue Th\'er\`ese-Lavoie-Roux, Montr\'eal, Queb\'ec, H2V 0B3, Canada}
\author[0000-0001-9754-2233]{Gerd Weigelt}
\affiliation{Max Planck Institute for Radio Astronomy, Auf dem H\"ugel 69, 53121 Bonn, Germany}
\author[0000-0002-8092-980X]{Peredur M. Williams}
\affiliation{Institute for Astronomy, Royal Observatory, Edinburgh EH9 3HJ, U.K.}

\author{Stefan Kraus}
\affiliation{Astrophysics Group, Department of Physics \& Astronomy, University of Exeter, Stocker Road, Exeter, EX4 4QL, UK}
\author{Jean-Baptiste Le Bouquin}
\affiliation{Institut de Planetologie et d'Astrophysique de Grenoble, Grenoble 38058, France}
\author{Narsireddy Anugu}
\affiliation{The CHARA Array of Georgia State University, Mount Wilson Observatory, Mount Wilson, CA 91203, USA}
\author{Sorabh Chhabra}
\affiliation{Astrophysics Group, Department of Physics \& Astronomy, University of Exeter, Stocker Road, Exeter, EX4 4QL, UK}
\author{Isabelle Codron}
\affiliation{Astrophysics Group, Department of Physics \& Astronomy, University of Exeter, Stocker Road, Exeter, EX4 4QL, UK}
\author{Jacob Ennis}
\affiliation{Astronomy Department, University of Michigan, Ann Arbor, MI 48109, USA}
\author{Tyler Gardner}
\affiliation{Astrophysics Group, Department of Physics \& Astronomy, University of Exeter, Stocker Road, Exeter, EX4 4QL, UK}
\author{Mayra Gutierrez}
\affiliation{Astronomy Department, University of Michigan, Ann Arbor, MI 48109, USA}
\author{Noura Ibrahim}
\affiliation{Astronomy Department, University of Michigan, Ann Arbor, MI 48109, USA}\author{Aaron Labdon}
\affiliation{European Southern Observatory, Casilla 19001, Santiago 19, Chile}
\author{Cyprien Lanthermann}
\affiliation{The CHARA Array of Georgia State University, Mount Wilson Observatory, Mount Wilson, CA 91203, USA}
\author{Benjamin R. Setterholm}
\affiliation{Astronomy Department, University of Michigan, Ann Arbor, MI 48109, USA}

\begin{abstract}

Classical Wolf-Rayet stars are the descendants of massive OB stars that have lost their hydrogen envelopes and are burning helium in their cores prior to exploding as type Ib/c supernovae. The mechanisms for losing their hydrogen envelopes are either through binary interactions or through strong stellar winds potentially coupled with episodic mass-loss. Amongst the bright classical WR stars, the binary system WR\,137 (HD\,192641; WC7d + O9e) is the subject of this paper. This binary is known to have a 13-year period and produces dust near periastron. Here we report on interferometry with the CHARA Array collected over a decade of time and providing the first visual orbit for the system. We combine these astrometric measurements with archival radial velocities to measure masses of the stars of $M_{\rm WR} = 9.5\pm3.4 M_\odot$ and $M_{\rm O} = 17.3\pm 1.9 M_\odot$ when we use the most recent \textit{Gaia} distance. These results are then compared to predicted dust distribution using these orbital elements, which match the observed imaging from \textit{JWST} as discussed recently by Lau et al. Furthermore, we compare the system to the BPASS models, finding that the WR star likely formed through stellar winds and not through binary interactions. However, the companion O star did likely accrete some material from the WR's mass-loss to provide the rotation seen today that drives its status as an Oe star.

\end{abstract}



\section{Introduction} \label{sec:intro}

Astronomical instrumentation has improved to facilitate precision measurements of exotic stellar systems that were not otherwise possible. Stars like classical Wolf-Rayet stars were historically modeled through the interpretation of single star evolution with episodic loss to allow for a 20--60 $M_\odot$ star to lose enough material to become a hydrogen-free, relatively compact object, a classical Wolf-Rayet star. This episodic mass-loss would happen in the short-lived and poorly understood phase of evolution represented by the luminous blue variable stars \citep[LBVs;][]{2006ApJ...645L..45S}. However, it has been shown in the last $\sim$decade that not only are most massive stars born in binary systems \citep{2013A&A...550A.107S,2014ApJS..215...15S}, but the vast majority of O stars, the progenitors of WR stars, are born in close enough systems that binary interactions such as Roche lobe overflow and merger scenarios can and will dominate the evolution of the O stars \citep{1998NewA....3..443V,2012Sci...337..444S}.

Because the binary interactions dominate the evolution of massive stars, it is critical to study some example systems in detail so that we can best constrain the masses and evolutionary pathways to create the systems we observe in the modern Universe. Traditionally, determination of binary-star masses requires core-eclipsing systems (i.e., systems where the eclipses occur when the projected stellar disks occult each other), but only a few classical nitrogen-rich Galactic WR stars have been observed in these configurations including: 
WR\,139 \citep[WN5 + O6III-V; $P = 4.0275$ d;][]{1941ApJ....93..202G, 1994ApJ...422..810M}, 
WR\,151 \citep[WN5 + O5;$P = 2.13$d][]{1948ApJ...108...56H, 1993ApJ...405..312L, 2009PASP..121..708H}, 
and WR\,155 \citep[WN6 + O9 II-Ib; $P = 1.64$d;][]{1944ApJ...100..242G, 1995ApJ...450..811M}. 
The system WR\,63 shows eclipses on a 4 d period, but this seems to come from an eclipsing binary of two O stars, with an effectively single WN7o star in a very long-period orbit around the O star binary if it is gravitationally bound \citep{2022MNRAS.516.1022C}. To date, no carbon-rich Wolf-Rayet systems have been found in eclipsing binaries, although there is one potential system being analyzed (Chene et al., in prep). We do not include other WR types in these numbers as the hydrogen-rich WNh stars are not an evolved class of objects and represent a massive extension to O stars.

In addition to the core-eclipsing systems, a second kind of photometric binary exists for the Wolf-Rayet stars. These binaries show what are called ``atmospheric eclipses" which allow for a determination of the binary inclination angle through modeling of the electron scattering through the Wolf-Rayet wind as the OB star passes behind the optically thick wind of the WR star. The light curve can then be modeled to provide an orbital inclination that is dependent on the mass-loss rate of the WR star \citep{1996AJ....112.2227L}. This technique has been used for dozens of Wolf-Rayet binaries, including the very massive and extreme main-sequence WNh stars in the R144 system \citep{2021A&A...650A.147S}. 

Two other methods exist that have been used with some reliability for deriving inclinations for short-period WR binaries, namely polarization variability and modeling of the colliding wind excess emission. The polarization method was first developed by \citet{1978A&A....68..415B} and uses Thomson scattering from the optically thin exterior part of the WR envelope as it is illuminated by any number of unpolarized point sources. This has worked for some O star binaries in this seminal paper and has been expanded to 10--20 WR binaries \citep[see literature review in ][]{2022ApJ...930...89F}. The collision of winds between the WR wind and an OB companion can result in a shocked gas with added emission, especially in the \ion{C}{3} $\lambda$5696 emission line of WC stars. This has been used for several WR binaries, especially in short, circular orbits as first demonstrated by \citet{2000MNRAS.318..402H} and \citet{2002MNRAS.335.1069H} for the three systems of WR\,42, WR\,79, and $\theta$ Mus. The extension of this technique for eccentric orbits is complicated by the fact that the emission line strength varies inversely proportionally to the distance between the stars in their orbit \citep{2011MNRAS.418....2F, 2017MNRAS.471.2715R}.

However, these systems that have been measured with photometric or polarimetric techniques are almost all guaranteed to have produced a Wolf-Rayet star through binary interactions as the periods are very short. \citet{2012Sci...337..444S} demonstrate that any system with a period of less than a few hundred days will interact, while \citet{1998NewA....3..443V} suggest interactions for any massive binary with a period up to 10 yr. An interesting observational challenge is to measure the precise orbits for longer period Wolf-Rayet binaries in order to compare these systems to binary evolution models to empirically determine where binary evolution dominates the production of WR stars compared to single star evolutionary paths. Therefore, it is important to obtain precise masses and orbital parameters for longer period WR binaries.

To date, only three classical Wolf-Rayet systems have well-established orbital inclinations without core-eclipses or atmospheric eclipses. These have been measured with interferometry. The closest WR binary, $\gamma^2$ Velorum, was the first binary star with a measured separation from an intensity interferometer \citep{1970MNRAS.148..103H}. The system, with spectral types of WC8 + O7.5III, is located only 330 pc away, and its orbit was first resolved by \citet{2007MNRAS.377..415N} with the Sydney University Stellar Interferometer. This orbit was later revisited by \citet{2017MNRAS.468.2655L} using the Very Large Telescope Interferometer and the AMBER instrument. Masses are now known to roughly 5\% precision from either of the orbital solutions. The binary WR\,140 (WC7 + O5.5I) has been a target of interferometry, first with the 3-telescope Infrared Optical Telescope Array \citep[IOTA; ][]{2004ApJ...602L..57M}, and with the first orbital solution presented by \citet{2011ApJ...742L...1M} who used data from the Center for High Angular Resolution Astronomy (CHARA) Array. This orbit has also been recently revisited with CHARA measurements by \citet{2021MNRAS.504.5221T}, who have provided masses for the highly eccentric ($e=0.8993\pm0.0013$), long period ($P = 2895.00\pm0.29$ d) binary to a precision better than 4\%. The final system with a visual interferometric orbit is the WN5o + O9I binary WR\,133, which was mapped with the CHARA Array over the 112 d orbit by \citet{2021ApJ...908L...3R} to obtain masses with a precision of 15\%, showing a need for additional work on this binary in the future. Lastly, we note here that \citet{2016MNRAS.461.4115R} also found that the long-period WR binaries WR\,137 and WR\,138 were resolved with CHARA, although their orbits are not yet fully mapped.

From the precise orbits of the two WC binaries $\gamma^2$ Vel and WR\,140, the evolutionary pathways to create the observed binaries have been explored. \citet{2009MNRAS.400L..20E} used binary evolution codes to show that the age of the $\gamma^2$ Vel system was 5.5 Myr rather than the previously assumed 3.5 Myr that was calculated from single-star evolutionary theory. This also provided a framework for rectifying the age of the massive WR binary with that of the Vela OB association. In the WR\,140 system, \citet{2021MNRAS.504.5221T} showed how the current day masses and hydrogen-free nature of the WC star could be obtained with binary evolution. Their model system shows how the modern day WR star has lost or transferred nearly 30 $M_\odot$ of material to reach its current measured mass of 10.3$M_\odot$. The high eccentricity of this system was explained in that tidally-enhanced mass transfer near periastron passages can cause perturbations in the orbit acting to increase the eccentricity rather than circularize the orbit as developed theoretically by \citet{2007ApJ...660.1624S,2007ApJ...667.1170S,2009ApJ...702.1387S,2010ApJ...724..546S}. In the $\gamma^2$ Vel system, \citet{2009MNRAS.400L..20E} explain the eccentric orbit by noting that the radiative envelopes of both components during mass-transfer dampen the tidal forces that would normally be used for circularizing the orbits. These systems show that binary interactions can be important even with very long periods for an evolved massive star, and thus are important in our interpretation of stellar populations.

In this paper, we focus on the carbon-rich Wolf-Rayet binary WR\,137 and present the first visual orbit for the system. This was one of the first three WR stars discovered through visual spectroscopy at the Paris Observatory \citep{1867CRAS...65..292W}. With moderate-resolution spectroscopy, \citet{Underhill62} suggested the system was a WC7+Be shell star binary but \citet{Massey} and \citet{Moffat} did not observe radial velocity variability, claiming that WR\,137 was therefore not a (close) binary system.

WR\,137 was observed to show a variable infrared brightness, which is associated with dust creation. The first outburst was observed by \citet{1985MNRAS.215P..23W} with a peak in mid-1984. Coupled with previous infrared data showing a decline in brightness, \citet{1985MNRAS.215P..23W} suggested that such eruptions could be periodic with a period of about 15~years. This prompted \citet{Annuk} to demonstrate that WR\,137 was a binary by presenting a first spectroscopic orbit, adopting a period of 4400 days from the IR data. Additional infrared photometry presented by \citet{2001MNRAS.324..156W} revealed another dust formation episode peaking in 1997, which then led to a period of 4765$\pm$50 days. Then, \citet{2005MNRAS.360..141L} used new and archival spectra to derive a double-lined spectroscopic orbit with a period of 4766$\pm$66 days, and a fairly low eccentricity of 0.178$\pm$0.042. The dust production peaks near periastron. 

\citet{2005MNRAS.360..141L} also presented evidence of a periodicity in the wind lines with a potential period of 0.83 d. This prompted \citet{2020MNRAS.497.4448S} to collect an intensive time-series of spectroscopic data during the summer of 2013 to search for these structures and determine if they were caused by co-rotating interaction regions in the wind. These were not found, but \citet{2020MNRAS.497.4448S} did show that the companion star should be considered an Oe star rather than a normal O star, as the O star lines tended to have fairly stable, double-peaked emission profiles reminiscent of the decretion disk profiles around Be stars, as was first suggested by \citet{Underhill62}. This equatorially-enhanced material could be necessary to form the dust in the way it is observed given the geometry of the dust plume imaged with aperture-masking interferometry with JWST/NIRISS \citep{2024ApJ...963..127L} and the nature of the dust properties shown with SOFIA spectroscopy by \citet{2023arXiv230811798P}.

This paper presents a visual orbit for WR\,137 and is organized as follows. Section \ref{sec:obs} presents our observations of the binary with the CHARA Array and discusses their reductions. Section \ref{sec:measure} describes our procedure for deriving differential astrometry from the infrared interferometry. We present the visual orbit in Section \ref{sec:orbit} and discuss our findings in light of binary evolution in Section \ref{sec:discuss}. We conclude this study in Section \ref{sec:jump-to-conclusions}. 

\section{Interferometric Observations} \label{sec:obs}
\subsection{CHARA Array observations} 

Our observations of WR\,137 were a continuation of the project begun by \citet{2016MNRAS.461.4115R} and utilized three instruments available at the CHARA Array \citep{2005ApJ...628..453T}. The CHARA Array is a $\Ydown$-shaped interferometric array of six 1-m telescopes with baselines ranging from 34 to 331 meters in length. We list the telescopes and instrument(s) used for each observing night in Table \ref{telescopes}. 

We re-reduced and analyzed data from the CLIMB beam combiner \citep{2013JAI.....240004T} taken over three nights on UT 2013 Aug 13-15 and originally published by \citet{2016MNRAS.461.4115R}. These measurements were taken with three telescopes at once, providing three baselines with which to measure squared visibilities and one measurement of closure phase (CP) per pointing. We also obtained new CLIMB observations in 2018. All measurements with CLIMB were taken with a calibrator star immediately preceding and following the observation in order to best calibrate the squared visibilities and closure phase. The calibrator stars used for all beam combiners along with their angular diameters and nights observed are listed in Table \ref{calibrators}. The CLIMB data were reduced with the pipeline developed by John D. Monnier; the general method is described in \citet{2011ApJ...742L...1M} and the extension to three beams is described in \citet{2018ApJ...855...44K}. During each epoch the CLIMB measurements were merged together over several consecutive nights and fit as a single binary position to improve the $(u,v)$ coverage. 

\begin{table*}
\centering
\caption{Telescopes and instruments used at the CHARA Array in our final analysis.
\label{telescopes}}
\begin{tabular}{l c c c}
\hline \hline
UT Night  &  Instrument    & Filter     & Telescopes \\
\hline
2013 August 13-15   &   CLIMB       &       $H$     &  S1-W1-E2 \\
2018 July 06-09     &   CLIMB       &       $H,K$     &  S2-W2-E2 \\
2019 July 01        &   MIRC-X      &       $H$     &  S2-W1-W2-E1-E2 \\
2019 July 02        &   MIRC-X      &       $H$     &  S1-S2-W1-W2-E1-E2 \\
2019 September 05   &   MIRC-X      &       $H$     &  S1-W1-E2 \\
2021 August 02      &   MIRC-X      &       $H$     &  S1-S2-W1-W2-E1-E2 \\
2021 October 22     &   MIRC-X      &       $H$     &  S1-S2-W1-W2-E1 \\
2021 October 22     &   MYSTIC      &       $K$     &  S1-S2-W1-W2-E1 \\
2022 July 19        &   MIRC-X      &       $H$     &  S1-S2-W1-W2-E1-E2 \\
2022 July 19        &   MYSTIC      &       $K$     &  S1-S2-W1-W2-E1-E2 \\
2022 August 23      &   MIRC-X      &       $H$     &  S1-S2-W1-W2-E1-E2 \\
2022 August 23      &   MYSTIC      &       $K$     &  S1-S2-W1-W2-E1-E2 \\
2023 June 03        &   MIRC-X      &       $H$     &  S1-S2-W1-W2-E1-E2 \\
2023 June 03        &   MYSTIC      &       $K$     &  S1-S2-W1-W2-E1-E2 \\
2023 August 14 (set \#1)      &   MIRC-X      &       $H$     &  S1-S2-W1-W2-E1-E2 \\
2023 August 14 (set \#1)      &   MYSTIC      &       $K$     &  S1-S2-W1-W2-E1-E2 \\
2023 August 14 (set \#2)      &   MIRC-X      &       $H$     &  S1-S2-W1-E1-E2 \\
2023 August 14 (set \#2)      &   MYSTIC      &       $K$     &  S1-S2-W1-E1-E2 \\
\hline \hline
\end{tabular}
\tablecomments{Additional CLIMB observations were obtained on UT 2018 June 05-06 and 2018 August 31, however, the $(u,v)$ coverage and data quality were not sufficient for measuring a reliable binary position. }
\end{table*}

\begin{table*}
\centering \rotate
\caption{Calibrator stars observed during the CLIMB, MIRC-X, and MYSTIC observations at the CHARA Array. A $\checkmark$ denotes the night this star was used as a calibrator. Calibrators found from the JMMC SearchCal database \citep{2006A&A...456..789B, 2011A&A...535A..53B}. \label{calibrators}}
\begin{tabular}{lccccccc}
\hline \hline
Calibrator star	&	$\theta_{{\rm UD},H}$ (mas)	&	$\theta_{{\rm UD},K}$ (mas)	&	2013 Aug 13	&	2013 Aug 14	&	2013 Aug 15	&	2019 Jul 01	&	2019 Jul 02	\\ \hline
HD\,178538	&	0.248715	&	0.249373	&		&		&		&	$\checkmark$	&	$\checkmark$	\\
HD\,191703	&	0.218459	&	0.219038	&	$\checkmark$	&	$\checkmark$	&	$\checkmark$	&	$\checkmark$	&	$\checkmark$	\\
HD\,192536	&	0.166190	&	0.166553	&		&	$\checkmark$	&	$\checkmark$	&		&		\\
HD\,201614	&	0.317421	&	0.318844	&		&		&		&	$\checkmark$	&		\\
HD\,197176	&	0.241453	&	0.242173	&		&		&		&	$\checkmark$	&	$\checkmark$	\\
HD\,192732	&	0.400280	&	0.402075	&		&		&		&		&		\\
HD\,192804	&	0.233558	&	0.234405	&		&	$\checkmark$	&		&		&		\\
\hline															
(continued)	&	2019 Sep 5	&	2021 Aug 02	&	2021 Oct 22	&	2022 Jul 19	&	2022 Aug 23	&	2023 Jun 03	&	2023 Aug 14	\\ \hline 
HD\,178538	&	$\checkmark$	&	$\checkmark$	&		&	$\checkmark$	&		&	$\checkmark$	&	$\checkmark$	\\
HD\,191703	&	$\checkmark$	&		&	$\checkmark$	&	$\checkmark$	&	$\checkmark$	&	$\checkmark$	&	$\checkmark$	\\
HD\,192536	&		&		&		&	$\checkmark$	&		&		&	$\checkmark$	\\
HD\,201614	&	$\checkmark$	&	$\checkmark$	&	$\checkmark$	&		&		&		&	$\checkmark$	\\
HD\,197176	&	$\checkmark$	&	$\checkmark$	&		&	$\checkmark$	&		&	$\checkmark$	&	$\checkmark$	\\
HD\,192732	&		&		&		&		&	$\checkmark$	&		&		\\
\hline \hline
\end{tabular}
\end{table*}

We also observed the system with the Michigan InfraRed Combiner - eXeter (MIRC-X) beam combiner \citep{2020AJ....160..158A}. This instrument combines up to all six telescopes at the CHARA Array and is an upgrade of the MIRC combiner \citep{2006SPIE.6268E..1PM} that allows for fainter targets to be observed with high precision. MIRC-X was used with the PRISM50 mode, allowing for 8 spectral channels across the $H-$band, with a spectral resolving power of $R\sim50$. Often the spectral channels at the edges of the $H-$band are rejected due to low signal-to-noise, meaning we end up with 6 spectral channels in each data-set. 

In August 2021, the CHARA Array commissioned a second six-telescope beam combiner, the Michigan Young Star Imager at CHARA \citep[MYSTIC;][]{2023JATIS...9b5006S}. MYSTIC observes in the $K$-band and operates simultaneously with MIRC-X. We used MYSTIC in PRISM49 mode, providing 11 spectral channels across the $K-$band with a spectral resolving power of $R\sim50$. Similarly to MIRC-X, the channels at the edges of the bandpass are often rejected, leaving us with nine useful wavelength channels across the $K-$band. 

The MIRC-X and MYSTIC data were reduced using the pipeline\footnote{https://gitlab.chara.gsu.edu/lebouquj/mircx\_pipeline} (version 1.3.3–1.3.5) developed by Jean-Baptiste Le Bouquin and the MIRC-X team, which splits each 10-minutes data sequence into four 2.5-minute bins. These reductions produce squared visibilities ($V^2$) for each baseline and CPs for each closed triangle of telescopes.
The use of 6 telescopes simultaneously allows for measurements of the squared visibility across 15 baselines with a simultaneous measurement of 20 different closure phases. With MIRC-X/MYSTIC, a single calibrated snapshot observation provides higher precision than was possible with CLIMB, as evidenced by the small astrometric errors presented in the orbit of the WN-type binary WR\,133 \citep{2021ApJ...908L...3R}. 

For each MIRC-X/MYSTIC night, we compared the calibrators against each other and found no evidence for binarity after visually inspecting the data. We applied wavelength correction factors by dividing the wavelengths in the MIRC-X OIFITS files by a factor of 1.0054 $\pm$ 0.0006 and those in the MYSTIC OIFITS files by 1.0067 $\pm$ 0.0007 \cite[][Monnier, priv. comm]{2022AJ....164..184G}.

\subsection{IOTA observations}

To complement the CHARA observations, we included the first spatially resolved measurement of WR 137 published by \citet{2010RMxAC..38...54R} that was obtained with the IOTA interferometer \citep{2003SPIE.4838...45T} at Mount Hopkins in Arizona. We added another previously unpublished IOTA observation obtained with the ICONIC3 combiner \citep{2003SPIE.4838.1099B} in the $H$-band from UT 2005 June 16. The projected baselines ranged from 18 to 38 meters. The data were reduced using the procedures described by \citet{2006ApJ...647..444M}.

\section{Astrometric Measurements} \label{sec:measure}

Our measurements from the calibrated interferometry were fitted for binary positions using a grid-search code\footnote{The code is available at https://www.chara.gsu.edu/analysis-software/binary-grid-search/} that has been used in multiple analyses including for the WR binaries discussed in \citet{2016MNRAS.461.4115R}, \citet{2021MNRAS.504.5221T}, and \citet{2021ApJ...908L...3R}. The code was originally described by \citet{2016AJ....152..213S} and uses both the measurements of fringe visibility and closure phase, which helps to remove a 180$^\circ$ ambiguity from the position angle. {For our analysis, we assume the stars are point sources as the stars should have angular diameters on the order of 20 $\upmu$as given the expected radii reported by \citet{2016MNRAS.461.4115R} and assuming a distance of $\sim 2$kpc \citep{2021AJ....161..147B}.} The fitting approach calculates the $\chi^2$ statistic between the data and a binary model for a large grid of separations in right ascension and declination. At each step in the grid, the IDL mpfit package \citep{2009ASPC..411..251M} is used to optimize the binary position and flux ratio between the two stars. The global minimum across the grid is selected as the best fit solution. We did a thorough search by varying the separations in increments of 0.5 mas across a range of $\pm20$ mas in both $\triangle$RA and $\triangle$DEC. For the CLIMB measurements on UT 2018 July 06-09, the binary solution with the lowest $\chi^2$ value was inconsistent with the orbital motion, however, an alternate solution within $\Delta \chi^2$ = 1.3 from the minimum was adopted as the final solution {as this was near the expected position given the MIRC-X and MYSTIC measurements}.

We present the binary separation ($\rho$) and position angle east of north (PA) in Table \ref{interferometry}. The error ellipses are defined by the major axis ($\sigma_{\rm major}$), minor axis $\sigma_{\rm minor}$, and position angle of the major axis ($\sigma_{\rm PA}$). Plots of the fits for each night of data are included in an online supplementary file. Informed by spectral modeling, we identified the brighter star in the $H$-band as the WR. Given the large distance to the binary, we assume two point sources in our analysis, so that the Oe disk and WR wind are considered all within these point sources. Given the system is about 6 times further away than $\gamma^2$ Vel where the similar stars have angular diameters of 0.47 and 0.22 mas for the O and WC stars respectively, the corresponding angular diameters for our MIRC-X observations would be less than 0.1 mas, which is unresolved by the CHARA Array. The measured visibilities of the binary did not rise all the way to a value of 1 at their peak, indicating that either one or both of the components could be marginally resolved or that there is excess flux outside of the interferometric field of view ($\sim$ 50 mas). Therefore, in addition to the relative flux contribution of the two components ($f_{\rm WR}$ and $f_{\rm O}$), we also included a visibility scaling factor to account for incoherent background flux ($f_{\rm incoherent}$). The contribution of the background ranged between 1\% to 18\% of the light in the $H$-band and 30\% to 50\% of the light in the $K$-band. The reduced $\chi^2$ for the binary fit was typically higher for MYSTIC data ($\chi^2_\nu = 10 - 30$) compared with MIRC-X ($\chi^2_\nu = 1 - 3$), indicating that there could be more complex extended emission in the $K$-band.

\begin{table*}
\centering
\caption{Interferometric measurements of the binary with the CHARA Array. 
\label{interferometry}}
\begin{tabular}{l c c c c c c c c c c c}
  \hline \hline
UT Date & HJD 	&	Filter	&	Separation	&	Position	&	$\sigma_{\rm major}$	&	$\sigma_{\rm minor}$	&	$\sigma_{\rm PA}$	&	$f_{\rm WR}$	&	$f_{\rm O}$	&	$f_{\rm incoherent}$ & Comb.	\\	
 & $-$2,400,000	 &		&	(mas)	&	Angle ($^\circ$)	&	(mas)	&	(mas)	&	($^\circ$)	&		&		&		\\	\hline
2005 June 16     & 53537.896    &	$H$	 &  9.43   &  114.21  & 0.60   & 0.17   & 119.2  &   0.451   &       0.549   &       \ldots  & I  \\
2005 July 7      & 53558.503    &	$H$	 &  9.80   &  115.00  & 0.60   & 0.22   & 115.0  &   0.448   &       0.552   &       \ldots  & I  \\
2013 August 14	 & 56518.827	&	$H$	 &  4.0209 &  132.011 & 0.0424 & 0.0330 & 136.41 &	0.618	&	0.382	&	\ldots  & C  \\	
2018 July 7      & 58307.325    &       $H,K$    & 10.4151 &  115.502 & 0.1815 & 0.0432 &  93.81 &   0.653   &       0.347   &       \ldots  & C  \\	
2019 July 1	 & 58665.725	&	$H$	 &  8.4677 &  112.738 & 0.0143 & 0.0124 &  98.06 &	0.547	&	0.426	&	0.027   & M	 \\	
2019 July 1	 & 58665.932	&	$H$	 &  8.4583 &  112.688 & 0.0112 & 0.0071 & 148.99 &	0.530	&	0.457	&	0.013   & M  \\	
2019 July 2      & 58666.767	&	$H$	 &  8.4523 &  112.807 & 0.0116 & 0.0112 & 171.65 &	0.505	&	0.433	&	0.061	& M  \\	
2019 July 2      & 58666.957	&	$H$	 &  8.4590 &  112.816 & 0.0108 & 0.0086 & 102.69 &	0.516	&	0.454	&	0.030	& M  \\	
2019 September 5 & 58731.902	&	$H$	 &  8.0081 &  112.370 & 0.0285 & 0.0141 &  51.12 &	0.471	&	0.428	&	0.101	& M  \\	
2021 August 2    & 59428.859	&	$H$	 &  1.4749 &   75.334 & 0.0045 & 0.0031 & 100.62 &	0.468	&	0.476	&	0.056	& M  \\	
2021 October 22  & 59509.642	&	$H$	 &  0.9940 &   37.459 & 0.0088 & 0.0053 & 109.55 &	0.472	&	0.445	&	0.083	& M  \\	
2021 October 22  & 59509.642	&	$K$	 &  0.9348 &   35.590 & 0.0204 & 0.0128 &  26.20 &	0.243	&	0.582	&	0.175	& Y  \\ 
2022 July 19     & 59779.761	&	$H$	 &  3.0377 &  312.477 & 0.0081 & 0.0029 & 120.09 &	0.464	&	0.393	&	0.142	& M  \\	
2022 July 19     & 59779.760	&	$K$	 &  3.0533 &  312.430 & 0.0098 & 0.0088 & 113.00 &	0.229	&	0.445	&	0.326	& Y  \\	
2022 August 23   & 59814.729	&	$H$	 &  3.3930 &  310.334 & 0.0049 & 0.0037 &  82.92 &	0.481	&	0.426	&	0.092	& M  \\	
2022 August 23   & 59814.729	&	$K$	 &  3.3936 &  310.357 & 0.0106 & 0.0086 & 127.23 &	0.188	&	0.427	&	0.385	& Y  \\	
2023 June 3      & 60098.790	&	$H$	 &  5.5326 &  300.915 & 0.0088 & 0.0037 & 110.42 &	0.486	&	0.362	&	0.152	& M  \\	
2023 June 3      & 60098.790	&	$K$	 &  5.5123 &  300.917 & 0.0177 & 0.0128 & 117.09 &	0.211	&	0.409	&	0.380	& Y  \\	
2023 August 14	 & 60170.711	&	$H$	 &  5.7794 &  299.396 & 0.0094 & 0.0086 &  99.39 &	0.494	&	0.330	&	0.176	& M  \\	
2023 August 14	 & 60170.711	&	$K$	 &  5.7687 &  299.363 & 0.0185 & 0.0112 & 111.21 &	0.170	&	0.384	&	0.446	& Y  \\	
2023 August 14	 & 60170.907	&	$H$	 &  5.7700 &  299.533 & 0.0096 & 0.0051 &  72.16 &	0.470	&	0.368	&	0.162	& M  \\	
2023 August 14	 & 60170.907	&	$K$	 &  5.7617 &  299.361 & 0.0269 & 0.0132 &  83.38 &	0.151	&	0.350	&	0.499	& Y  \\	\hline \hline
\end{tabular}
\tablecomments{Combiner codes: I = IOTA IONIC, C = CHARA CLIMB, M = CHARA MIRC-X, Y = CHARA MYSTIC}
\end{table*}

\section{The Visual Orbit of WR 137} \label{sec:orbit}

With our multi-epoch astrometric measurements in hand, we began by fitting a visual orbit. We utilize the software tools\footnote{http://www.chara.gsu.edu/analysis-software/orbfit-lib} described by \citet{2016AJ....152..213S}. We first fit all measurements from CHARA, including CLIMB, MIRC-X, and MYSTIC measurements, and IOTA with a purely visual orbit. The starting parameters were based on the radial velocity orbit of WR\,137 presented by \citet{2005MNRAS.360..141L}, which included the period $P$, eccentricity $e$, argument for periastron $\omega_{\rm WR}$, and the time of periastron passage $T$. The visual orbit solves also for the angular semi-major axis $a$, inclination $i$, and position angle of the line of nodes $\Omega$. We scaled the uncertainties on the CHARA measurements by a factor of 2.0 to force the reduced $\chi^2$ statistic to a value of unity as the errors were likely underestimated by the grid search routine. We used the published uncertainty for the IOTA measurement reported by \citet{2010RMxAC..38...54R} and scaled the uncertainty on the newly published IOTA measurement so that the major axis of the error ellipse for both measurements was the same. The final scaled uncertainties for all of the interferometric measurements are presented in Table \ref{interferometry}. 

\begin{figure}
    \centering
    \includegraphics[angle=0, width=3.5in]{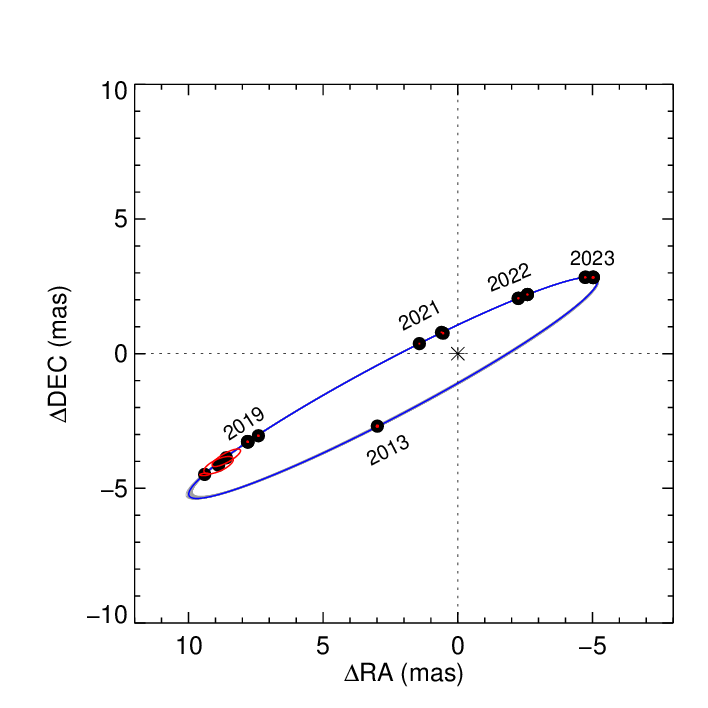}
    \includegraphics[angle=0, width=3.5in]{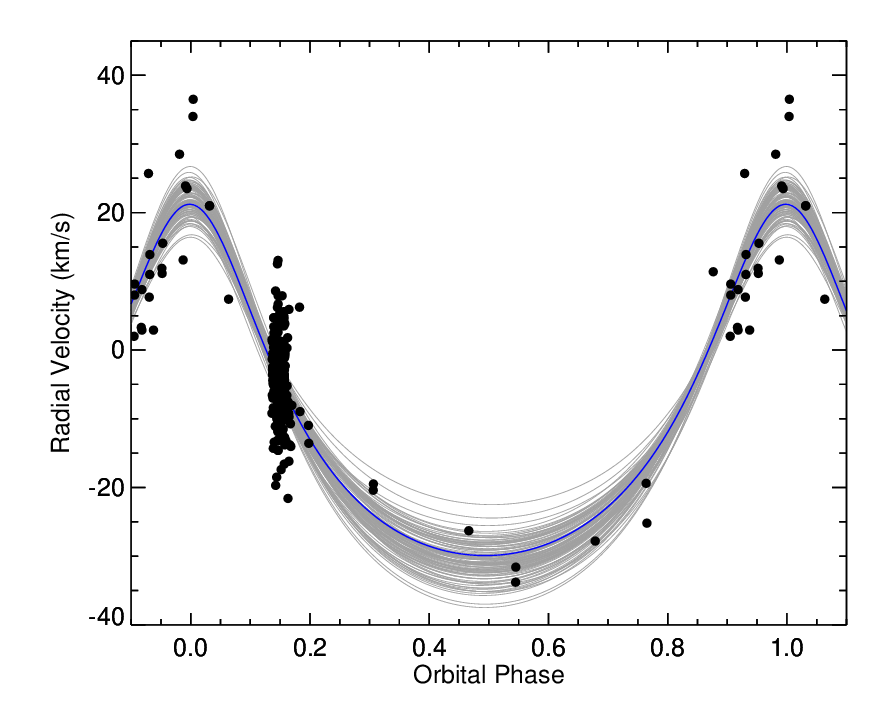}
    \caption{The orbital solution for the CHARA measurements presented in this analysis along with the measurements of the WR star's radial velocities published by \citet{2005MNRAS.360..141L}. For the visual orbit, the model is shown in blue with the measurements shown as black dots. The red ellipses show the measurement errors of the interferometry (typically 10--50 $\upmu$as). We have noted the year for observations with the CHARA Array including CLIMB (2013) and the MIRC-X combiner (2019--2023), with the MYSTIC observations taken simultaneously blending in with the MIRC-X observations. The IOTA observations were at a similar phase as our 2019 observations, which are shown with their larger error ellipses. The visual orbit shows the path of the WC star around the O star. We used the data from \citet{2005MNRAS.360..141L} for the radial velocities, who did not include formal errors on the points but rather stated that the error for each measurement was $\sigma \sim 5-10$ km s$^{-1}$. We also show the potential orbital solutions in grey with the adopted solution in blue. Note that the uncertainty in the visual orbit is almost negligible, while work remains to improve upon the spectroscopic orbit. }
    \label{fig:orbit}
\end{figure}

With a realistic handle of the errors in the interferometric measurements, we then used the spectroscopic measurements of the WR star published by \citet{2005MNRAS.360..141L} to do a combined fit of a spectroscopic and visual orbit simultaneously. This combined orbit is shown in Fig.~\ref{fig:orbit} and the orbital elements are given in Table \ref{orbital elements}. We also attempted to solve for an orbit with the velocities of the O star provided by \citet{2005MNRAS.360..141L}, but these velocities were unable to be fit with a value of $\omega_{\rm O}$ that was opposite that of the WR star or with a meaningful error on the semi-amplitude. We note that the analysis of \citet{2005MNRAS.360..141L} fit velocities with a Gaussian for the \ion{He}{1} $\lambda$5876 line. \citet{2020MNRAS.497.4448S} showed that the O star was an Oe star that has emission components in the \ion{He}{1} lines, so the strong line at 5876\AA\ was likely contaminated by disk emission. Thus, these measurements represent more of the disk geometry than orbital motion and we did not incorporate them. {We computed uncertainties in the orbital parameters through a Monte Carlo bootstrap approach, where we randomly selected positions and radial velocities from the sample of measured values with repetition. We then randomly varied the sample of measured values within their uncertainties and re-fit the orbit. We repeated this process 10,000 times and adopted uncertainties from the standard deviation of the bootstrap distributions. Corner plots showing correlations between the orbital parameters are shown in the Appendix Fig.~\ref{fig:corner}. Our resulting combined orbit of the WR and O star is slightly longer than the period for the radial velocity orbit reported by \citet{2005MNRAS.360..141L} or the infrared light curve period reported by \citet{2001MNRAS.324..156W} but is still within their errors. Our use of interferometry shows that the system also has a higher eccentricity than the orbit reported by \citet{2005MNRAS.360..141L}. The sampling of the radial velocity data could have also led to the lower eccentricity in the previous radial velocity orbit. }

\begin{table*}
\centering
\caption{Orbital Elements 
\label{orbital elements}}
\begin{tabular}{l c c c}
\hline \hline
\multicolumn{4}{c}{Measured Quantities} \\ \hline
Orbital Element	&	\multicolumn{3}{c}{Value}		 \\		
\hline				
$P$(d)	&	\multicolumn{3}{c}{		4786.5   $\pm$ 12.6	 }\\						
$P$(yr)	&	\multicolumn{3}{c}{		13.105	$\pm$	0.034  }\\						
$T$ (JD)	&	\multicolumn{3}{c}{		2,460,258.6  $\pm$ 7.3  }\\	
$T$ (yr)	&	\multicolumn{3}{c}{		2023.857	$\pm$	0.020	}\\	

$e$	&	\multicolumn{3}{c}{		0.3162   $\pm$ 0.0023}\\						
$a$(mas):       	&	\multicolumn{3}{c}{		8.575   $\pm$ 0.020  }\\						
$i$	&	\multicolumn{3}{c}{		97.138   $\pm$ 0.063  }\\						
$\Omega$ ($^\circ$)	&	\multicolumn{3}{c}{		117.934  $\pm$ 0.039	}\\						
$\omega_{\rm WR}$  ($^\circ$)        	&	\multicolumn{3}{c}{		361.24   $\pm$ 0.99	}\\						
$K_1$ (km s$^{-1}$)       	&	\multicolumn{3}{c}{		25.6     $\pm$ 2.2	}\\						
$\gamma$ (km s$^{-1}$)    	&	\multicolumn{3}{c}{		-12.4   $\pm$ 1.2}\\						
$\chi^2$    	&	\multicolumn{3}{c}{		284.4			}\\						
$\chi^2_{\rm red}$    	&	\multicolumn{3}{c}{		0.86			}\\						
\hline													
\multicolumn{4}{c}{Derived Quantities} \\ \hline
Quantity	&	Fit, $K_2=6.3$ km s$^{-1}$	&	Fit, d=2.11 kpc	&		Fit, d=1.94 kpc	\\ 	\hline 															
$M_{\rm WR}$ ($M_\odot$)	&	2.65	$\pm$	2.36	&	14.11	$\pm$	9.17	&	9.49	$\pm$	3.41	 \\
$M_{\rm O}$ ($M_\odot$)	    &	11.11	$\pm$	3.11	&	20.58	$\pm$	3.74	&	17.34	$\pm$	1.91	 \\
a1(AU)	                     &	10.75	$\pm$	0.54	&	10.76	$\pm$	0.91	&	10.76	$\pm$	0.91	 \\
a2(AU)	                     &	2.57	$\pm$	1.64	&	7.37	$\pm$	1.72	&	5.88	$\pm$	1.08	 \\
d(pc)	                      &	1553	$\pm$	202	&	2114	$\pm$	160	&	1941	$\pm$	71	 \\
Parallax (mas)	             &	0.644	$\pm$	0.084	&	0.473	$\pm$	0.038	&	0.515	$\pm$	0.018	 \\
Reference (d)               & Derived  & \citet{2020MNRAS.493.1512R} & \citet{2021AJ....161..147B} \\
\hline \hline
\end{tabular}    
\tablecomments{$\omega_{\rm WR}$ + $\omega_{\rm O}$ = 180$^\circ$.}
\end{table*}

With a visual and spectroscopic orbit, we can calculate masses for the component stars. If we assume that the value of $K_2$ presented by \citet{2005MNRAS.360..141L} was the actual semi-amplitude of the O star, $6.1\pm1.3$ km s$^{-1}$, then we can calculate an orbital parallax for the system and derive individual masses. This is shown in the bottom portion of Table \ref{orbital elements} in the column labeled $K_2 = 6.3$ km s$^{-1}$. This provides an implausibly small mass for the WR star of only 2.7$\pm$2.4 $M_\odot$. The O star mass also seems low with $M_{\rm O} = 11.1\pm3.1 M_\odot$. This is much lower than expected values of an O9-O9.5V star from the spectroscopic calibrations of \citet{2005A&A...436.1049M}, who found the late-O dwarfs should have masses of 16.6--18 $M_\odot$ as expected by the Oe nature of the companion \citep{2020MNRAS.497.4448S}. The distance of WR\,137 using this method is about 1.5 kpc, which is also smaller than expected \citep{2020MNRAS.493.1512R}.

In the absence of a double-lined binary, we can still calculate masses if we know the distance to the binary system. To that end, we used two distances from the literature. The first, 2.1 kpc, is taken from \citet{2020MNRAS.493.1512R} who examined the Galactic population of WR stars and examined their distances based on the \textit{Gaia} mission DR2 distances. This provides higher masses for the stars, namely 12.9 $M_\odot$ for the WR star and 20.1 $M_\odot$ for the O star. The O star mass is now higher than expected, with the mass value for the WR star higher than the two other dynamically-measured WC star masses, which were $\gamma^2$ Velorum \citep[9 $M_\odot$; ][]{2017MNRAS.468.2655L, 2007MNRAS.377..415N} and WR\,140 \citep[10.3 $M_\odot$; ][]{2021MNRAS.504.5221T, 2011ApJ...742L...1M}. We note that the first measurement of the mass of WR\,140 by \citet{2011ApJ...742L...1M} was 14.9 $M_\odot$, but the newer orbit of \citet{2021MNRAS.504.5221T} incorporated a much better data set along with archival measurements. 

Finally, we used the distance derived by a Bayesian treatment of the parallaxes measured in the early data-release 3 from \textit{Gaia} (EDR3). This was presented by \citet{2021AJ....161..147B} and the distance to WR\,137 is 1.94 kpc, with an error of less than 100 pc. This provides a WR mass similar to the other WC stars, 8.6 $M_\odot$. The O star mass is in line with the spectroscopic predictions for its spectral type of 16.9 $M_\odot$, so we adopt this solution in our interpretation of the binary, but we present all three sets of derived quantities for the orbit in Table \ref{orbital elements} for completeness. Beyond masses, we also include the semi-major axis in AU for both stars along with distance and parallax for each solution. 
{While both distance estimates \citep{2020MNRAS.493.1512R, 2021AJ....161..147B} have advantages and disadvantages to their adoption, the distances are compatible with each other as are the masses of the stars within the derived errors. We suspect that the distance from the EDR3 of \textit{Gaia} \citet{2021AJ....161..147B} is the more appropriate solution, but we leave both for completeness in our discussion. For the remaining analysis, we will consider the \textit{Gaia} EDR3 distance-based solution for stellar masses.}

\section{Discussion} \label{sec:discuss}

\subsection{The impact of the visual orbit and CHARA measurements on the dust formation for WR 137}

With a visual and spectroscopic orbit, we can now consider how the dust production is impacted with the geometry of the orbit. Recently, \citet{2022NatAs...6.1308L} presented \textit{JWST}+MIRI imaging of the prototype of the dust-forming WC binaries, WR\,140. These images were compared to a geometric model from the colliding wind geometry both by \citet{2022NatAs...6.1308L} and in more detail by \citet{2022Natur.610..269H}. With the orbit well constrained both spectroscopically and interferometrically \citep[most recently by ][]{2021MNRAS.504.5221T}, the only free parameters in this geometric model are the dust expansion velocities and the phases/distances from the star where the dust can condense. The agreement between these models and the observations provides us evidence of how to model the dust production in these systems. As the only other WR binaries with the same level of orbital precision in the literature are WR\,133 \citep[WN5o + O9I; ][]{2021ApJ...908L...3R} and $\gamma^2$ Vel \citep[WC8 + O7.5III; ][]{2017MNRAS.468.2655L} do not form dust, WR\,137 now offers us an opportunity to truly test these geometric models.

We used our orbital parameters for WR\,137 to calculate a geometric model for predicted dust emission from the system. There have been two studies that have spatially resolved the dust surrounding the WR\,137 system. The first of these used the near-infrared NICMOS camera on the \textit{Hubble Space Telescope} using a comparison of the point-spread-function of WR\,137 during a dust creation event and that of a simpler WR star, WR\,138. This imaging reported by \citet{1999ApJ...522..433M} suggested a dust plume that pointed slightly south of west from the central source at a similar orbital phase as the \textit{JWST}+NIRISS aperture-masking interferometry observations that were recently obtained and presented by \citet{2024ApJ...963..127L}. 

In recent years, work on dusty WC binaries such as that reported by \citet{2020ApJ...900..190L} for WR\,112, has shown that the orbital elements could potentially be inferred by examining the morphology changes of the dust cloud with time. Repeatability of the geometry provides a means to measure the period, and then geometric models based on the orbital elements, the wind momentum balance in the system, and the phases at which dust is formed can reproduce the overall morphology inferred from the infrared imaging. This has been used to explain the ground-based imaging of WR\,112 \citep{2020ApJ...900..190L} and Apep \citep{2020MNRAS.498.5604H}, along with \textit{JWST} imaging of WR\,140 \citep{2022Natur.610..269H}.

\citet{2016MNRAS.461.4115R} presented a spectroscopic model of the binary system WR\,137 and thus provided exquisite constraints on the mass-loss momentum between the two stars. Thus, with our visual orbital elements measured with CHARA and the archival spectroscopic measurements of \citet{2005MNRAS.360..141L}, we only have to provide constraints on the orbital phases when dust is produced. The results of the models were presented with a comparison to the \textit{JWST} + NIRISS aperture masking interferometry images by \citet{2024ApJ...963..127L}. The results are in strong agreement between the modeling of the geometry of the dust and the fundamental orbital parameters measured neglecting any uncertainties due to the decretion disk around the O star in WR\,137 which seems to make a thinner dust plume in this system. We can therefore see that with great agreement for both WCd systems with established visual and spectroscopic orbits, namely WR\,137 in this paper and WR\,140 \citep{2022Natur.610..269H, 2021MNRAS.504.5221T}, dust geometry can be used as a means to infer orbital elements in the absense of long-term measurments of radial velocities or astrometry.


In addition to these geometric models for the dust around WR\,137, we also had to fit a flux from the large-scale dust distribution around the binary, often called incoherent flux in interferometry, to the MIRC-X ($H$-band) and MYSTIC ($K$-band) measurements to obtain the astrometry of the system. These flux measurements are needed in this system and were also used in the fits for WR\,140 \citet{2021MNRAS.504.5221T}. However, in the WN-type binary WR\,133, this incoherent flux was not needed \citep{2021ApJ...908L...3R}. The overall field of view for these instruments is on the order of tens of milliarcseconds, so this incoherent flux seen in the dust making systems WR\,140 and WR\,137 could easily be attributed to the large-scale dust emission that is within the field of view of the interferometer. The interferometer will not likely be able to spatially resolve a large-scale dust structure as it extends well beyond the nominal imaging field for the instruments. 

\begin{figure}
    \centering
    \includegraphics[angle=0, width=6in]{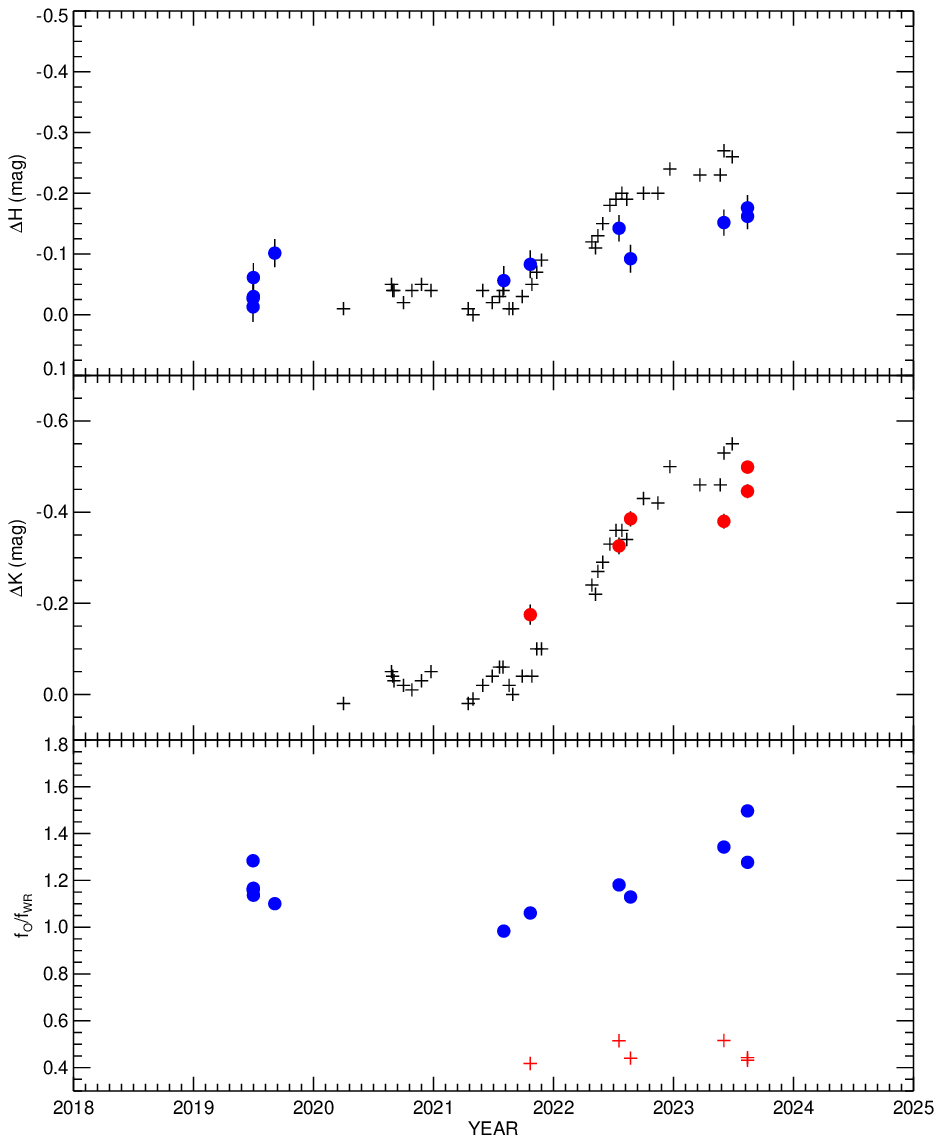}
    \caption{The recent infrared light curve of WR\,137 from \citet{2023arXiv230811798P} is shown for the $H$- and $K$-bands as $+$ symbols, shifted to $\triangle$mag being around 0 for the times prior to the current dust formation episode (2020--late 2021). The blue/red circles represent the incoherent flux fitted in the CHARA measurements in the $H/K$-band. In the bottom panel, we show the ratio of the WR and O star fluxes, which are nearly constant within our errors. }
    \label{fig:incoherent}
\end{figure}

In Fig.~\ref{fig:incoherent}, we show the infrared light curve from \citet{2023arXiv230811798P} along with the incoherent flux measured from our interferometry in the $H$- and $K$-bands. While the interferometric points likely have a larger uncertainty of several percent, we see the $H$-band incoherent flux was usually measured to be a few percent in our data from 2019 through mid 2021. More recently, the MIRC-X measurements in $H$-band have shown an incoherent contribution of about 15\%, and the light curve shows an excess of about 0.25 mag. While we do not have the same sort of time coverage in the $K$-band with MYSTIC, we see an average incoherent flux around 30--50\% with the $K$-band light curve excess around 0.5 mag and increasing after the first observation. Thus, we expect that the CHARA Array is seeing the dust emission, but the interferometer is unable to image this too extended dust emission. In order to show consistency within our binary fits, we also show the ratio of the flux of the WR and O star in both bandpasses where the ratio is almost constant within the errors of the flux estimates for each component. The first MYSTIC observation may be an outlier, but the instrument was still new at that time so the calibration frames may not have been adequate for precise measurements. We also note that the measurements of the flux of the WR star in the $K-$band are dominated by the large \ion{C}{4} complex at 2.08$\upmu$m along with \ion{C}{3}$+$\ion{He}{1} at 2.11$\upmu$m \citep{1997ApJ...486..420F}.

\subsection{The evolutionary status of WR 137}

We can use the observational parameters of WR\,137 to try and understand its evolutionary history and future by comparing its parameters to binary evolution models from the Binary Population And Spectral Synthesis (BPASS) code, v2.2.1 models, as described in detail in \citet{2017PASA...34...58E} and \citet{2018MNRAS.479...75S}. We follow the fitting method in \citet{2009MNRAS.400L..20E} and \citet{2011MNRAS.411..235E}. We use the $UBVJHK$ magnitudes taken from\citet{2002yCat.2237....0D}  and  \citet{2003yCat.2246....0C}. To estimate the extinction, we take the $V-$band magnitude from the BPASS model for each time-step and compare it to the observed magnitude. If the model $V-$band flux is higher than observed, we use the difference to calculate the current value of $A_V$ . If the model flux is less than observed, we assume zero extinction. We then modify the rest of the model time-step magnitudes with this derived extinction before determining how well that model fits. The current measurement of $A_V$ is 1.85 \citep[e.g.,][]{1988A&A...199..217V}. We then also require that the model must have a primary star that is now hydrogen-free, have carbon and oxygen mass fractions that are higher than the nitrogen mass fraction and that the masses of the components and their separation match the observed values that we determined here.

The one caveat in our fitting is that the BPASS models assume circular orbits; however, as found by \citet{2002MNRAS.329..897H}, stars in orbits with the same semi-latus rectum, or same angular momentum, evolve in similar pathways independent of their eccentricity. A similar assumption was made in \citet{2009MNRAS.400L..20E} for $\gamma^2$ Velorum. We note that a more realistic model would require including the eccentricity. WR\,137’s moderate eccentricity could be a system where specific modeling of the interactions may lead to interesting findings on how and if the stars in long-period eccentric binaries can interact \citep[see][for examples]{2016ApJ...825...70D, 2016ApJ...825...71D}.

\begin{figure}
    \centering
    \includegraphics[width=\columnwidth]{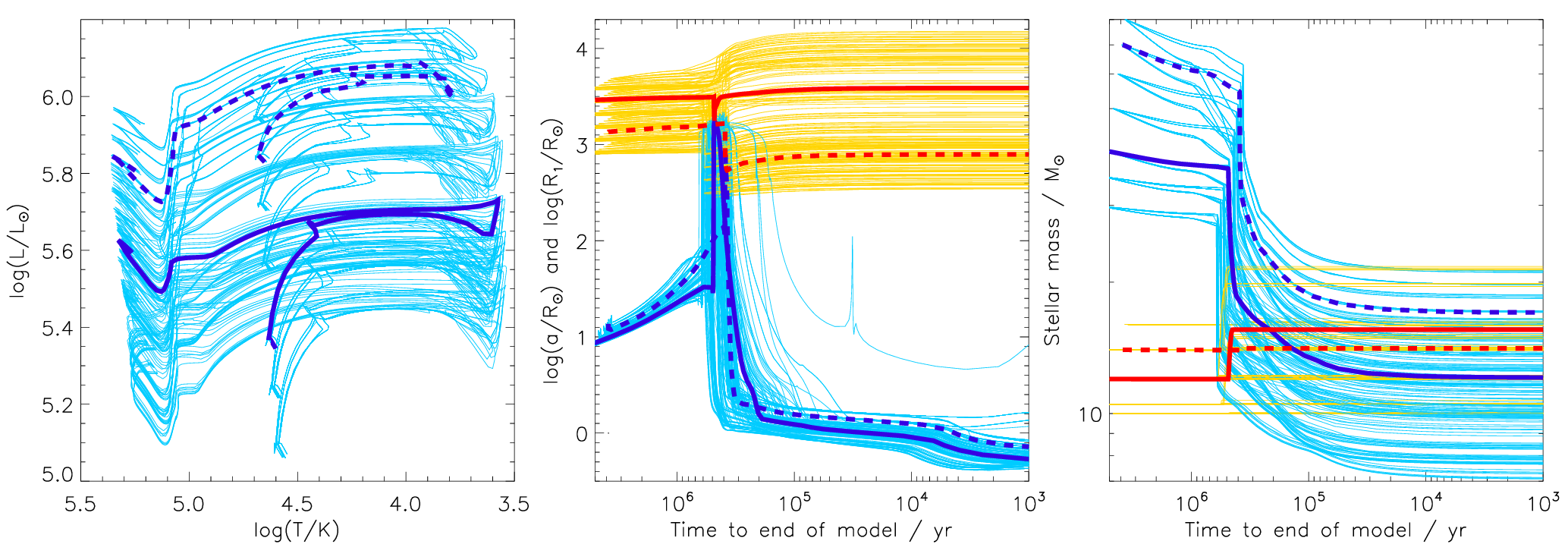}
    \caption{Different aspects of evolution of the WR\,137 system are shown in these three panels. {The blue (WR star) and red (O star)} bold lines represent the model with the best matching initial parameters with thinner lined models that are within the 1$\sigma$ uncertainties in initial mass, initial mass ratio, initial period and initial metallicity. The mean model is shown as a dashed line while the mode model is shown as a thick line. In the left panel, we show the Hertzsprung-Russell diagram for the past and future evolution of the WR star. In the central panel, we show the primary radius in light/dark blue and the orbital separation in yellow/red. In the right panel we show the mass of the primary in light/dark blue and the mass of the secondary in yellow/red.}
    \label{fig:evolution}
\end{figure}

We find a broad range of models that match WR\,137, although we are able to identify two possible evolutionary pathways among the large range of models that fit. First, the mean fit gives a pathway at solar metallicity (Z=0.017$\pm$0.007) and initial masses for the stars of 69$\pm$21$M_\odot$ for the current-day WC star, and 15.5$\pm$2.5$M_\odot$ for the Oe star. The initial periods are in the range of log P[d] = 2.8$\pm$0.4,. The derived extinction for this pathway is 1.58$\pm$0.06. 
In this pathway, we find that many models fill their Roche lobes so that some mass is transferred to the companion. While for some of the models, this does not occur so one might think that no mass transfer occurs. However, given that these models nearly fill their Roche lobes, mass transfer might still occur. \citet{2021PASA...38...56H} find that when a star fills $>$80\% of the volume of it's Roche lobe \textit{wind fed Roche lobe overflow} can occur where the stellar wind is focused such that mass is accreted by the companion star which may explain the observed features of the Oe star in the WR\,137 binary.

In this pathway we find that while many models fill their Roche lobes for some the primary star does not. Thus WR\,137 here has been formed primarily due to the stellar wind mass loss.

The second pathway is best represented by the mode of the fitting models and is preferred at lower metallicities of half solar (Z=0.010$\pm$0.009) at initial masses of 35$\pm$21~M$_{\odot}$ for the WC star and 10.5$\pm$2.5$M_\odot$ for the Oe star. In this fit, the initial periods are now in the range of $\log$P[d] = 3.4$\pm$0.4. With a lower $A_V$=1.54$\pm$0.06 and an older age of log(age/yrs)=6.71$\pm$0.06. The initial periods can be greater because to reach the current mass of the WC star less mass must be lost so less orbital widening is required. Here, while for some of the models again the WC star forms due to stellar winds alone there are several models where a short period of mass transfer occurs. We show example fits in Fig~\ref{fig:evolution}, with the closest matching model highlighted.

\begin{table}[]
    \centering
    \begin{tabular}{l c c}
    \hline \hline 
      Parameter & Mean Fit & Mode Fit \\ \hline 
      $M_1$ (current WC; $M_\odot$)    & 69$\pm$21  & 35$\pm$21 \\
      $M_2$ (current Oe; $M_\odot$)    & 15.5$\pm$2.5 & 10.5$\pm$2.5 \\
      $\log P_{\rm initial}$    & 2.84$\pm$0.37 & 3.4$\pm$0.37 \\
      $\log ({\rm age [yr]})$ & 6.60$\pm$0.08 & 6.71$\pm$0.08 \\
      $A_V$ & 1.58$\pm$0.06 & 1.54$\pm$0.06 \\
      $Z$ & 0.017$\pm$0.007 & 0.010$\pm$0.009 \\ \hline \hline
    \end{tabular}
    \caption{The median and mode models for the current day WR\,137 system from BPASS v2.2.1. The evolutionary pathways are shown in Fig.~\ref{fig:evolution}.}
    \label{models}
\end{table}

Both pathways are possible and are compared in Table \ref{models} and shown in Fig.~\ref{fig:evolution}, the former has more matching models but less mass transfer to the secondary. The latter has more occurrence of significant mass transfer onto the companion. It is the amount of mass transfer that makes us prefer the second pathway due to the Oe nature of the companion. This suggests significant mass and angular momentum transfer must have occurred. However it is still possible that the more massive pathway could have led to at least some mass transfer, even if the primary did not fill it's Roche lobe. It has been suggested that when a star comes close to filling it's Roche lobe then, "wind-fed Roche lobe overflow" can occur. Where the stellar wind can be focused so that rather than being lost from the system some mass is transferred to the companion \citep[see][]{2007ASPC..372..397M,2021PASA...38...56H}. For our wider models where stellar winds alone formed the WC star this process must have occurred to create the Oe star.

Our results imply that, unlike the models for WR\,140 \citep{2021MNRAS.504.5221T} or $\gamma^2$ Vel \citep{2009MNRAS.400L..20E}, interactions between the two stars may have been weak or non-existent in most cases as the system began in a wide orbit and remains in such a wide orbit because of the mass loss history needed to get the orbit observed today. Our results show that WR\,137 contains a WR star that could have formed as a single star, but the companion is likely to have gained mass from either a weak interaction or by accreting wind material. These results assume circular orbits which do not reflect the current system, but the models give us an indication of the evolutionary history since stellar systems with the same angular momentum should evolve in similar ways \citep{2002MNRAS.329..897H}.

\section{Conclusions} \label{sec:jump-to-conclusions}

We have presented the first visual orbit of the dust-making binary WR\,137, providing measured masses of the component stars of $M_{\rm WR} = 8.61\pm3.05 M_\odot$ and $M_{\rm O} = 16.92\pm 1.46 M_\odot$. This places the Wolf-Rayet star being similar mass of WC star in WR\,140 that has a mass of $M_{\rm WR} = 10.31\pm 0.45 M_\odot$, with the same spectral types of the WR stars. The O9 star in the system has a mass of $16.92\pm1.46 M_\odot$, which is not too different than the mass of the O9III star in the $\iota$ Ori system that has a mass of 23.2$M_\odot$ from the binary analysis of \citet{2017MNRAS.467.2494P}. 

In addition to these fundamental measurements for the binary, the reproduction of the dust geometry for both WR\,140 and WR\,137 implies that these geometric dust models for WR binaries can yield reasonably well-constrained orbits. For some of these binaries, the orbital periods, as inferred from the geometric variations and potential periods are on the order of several decades. Such an orbit could be impractical to measure spectroscopically both due to the needed telescope allocations as well as the low amplitudes of the component stars, and while in principle they could be measured with interferometry, that also provides a challenge. The incoherent flux from the dust creates large challenges for interferometric detection of the two stars in the infrared, while this dust also attenuates the source forbidding optical work. Thus, the verification of the dust model for WR\,140 and WR\,137 provides a framework with which to infer the orbital architectures and thus fundamental properties of WCd stars.

\clearpage

\begin{acknowledgments}

This work is based upon observations obtained with the Georgia State University Center for High Angular Resolution Astronomy Array at Mount Wilson Observatory. The CHARA Array is supported by the National Science Foundation under Grant No. AST-1636624 and AST-2034336.  Institutional support has been provided from the GSU College of Arts and Sciences and the GSU Office of the Vice President for Research and Economic Development. Time at the CHARA Array was granted through the NOIRLab community access program (NOIRLab PropIDs: 2017B-0088, 2021B-0159, and 2023A-452855; PI: N. Richardson). This research has made use of the Jean-Marie Mariotti Center Aspro and SearchCal services. We thank Claire Davies, Theo ten Brummelaar, Dan Mortimer for past contributions to MIRC-X and MYSTIC that made this work possible.

NDR is grateful for support from the Cottrell Scholar Award \#CS-CSA-2023-143 sponsored by the Research Corporation for Science Advancement. SK acknowledges funding for MIRC-X received funding from the European Research Council (ERC) under the European Union's Horizon 2020 research and innovation programme (Starting Grant No. 639889 and Consolidated Grant No. 101003096). JDM acknowledges funding for the development of MIRC-X (NASA-XRP NNX16AD43G, NSF-AST 1909165) and MYSTIC (NSF-ATI 1506540, NSF-AST 1909165).

\end{acknowledgments}

%

\vspace{5mm}
\facilities{CHARA}


\software{astropy \citep{2013A&A...558A..33A,2018AJ....156..123A},  
          }



 \appendix
 In this appendix, we include a figure set showing the interferometric fits to the observations to obtain the separation and position angle of the binary at these epochs along with a corner plot showing the uncertainties and their correlations in the orbital fit. Note that for the arXiv version of the paper, we have only included one example plot of the fitted binary positions.

 \begin{figure}
     \centering
     \includegraphics[width=0.85\linewidth]{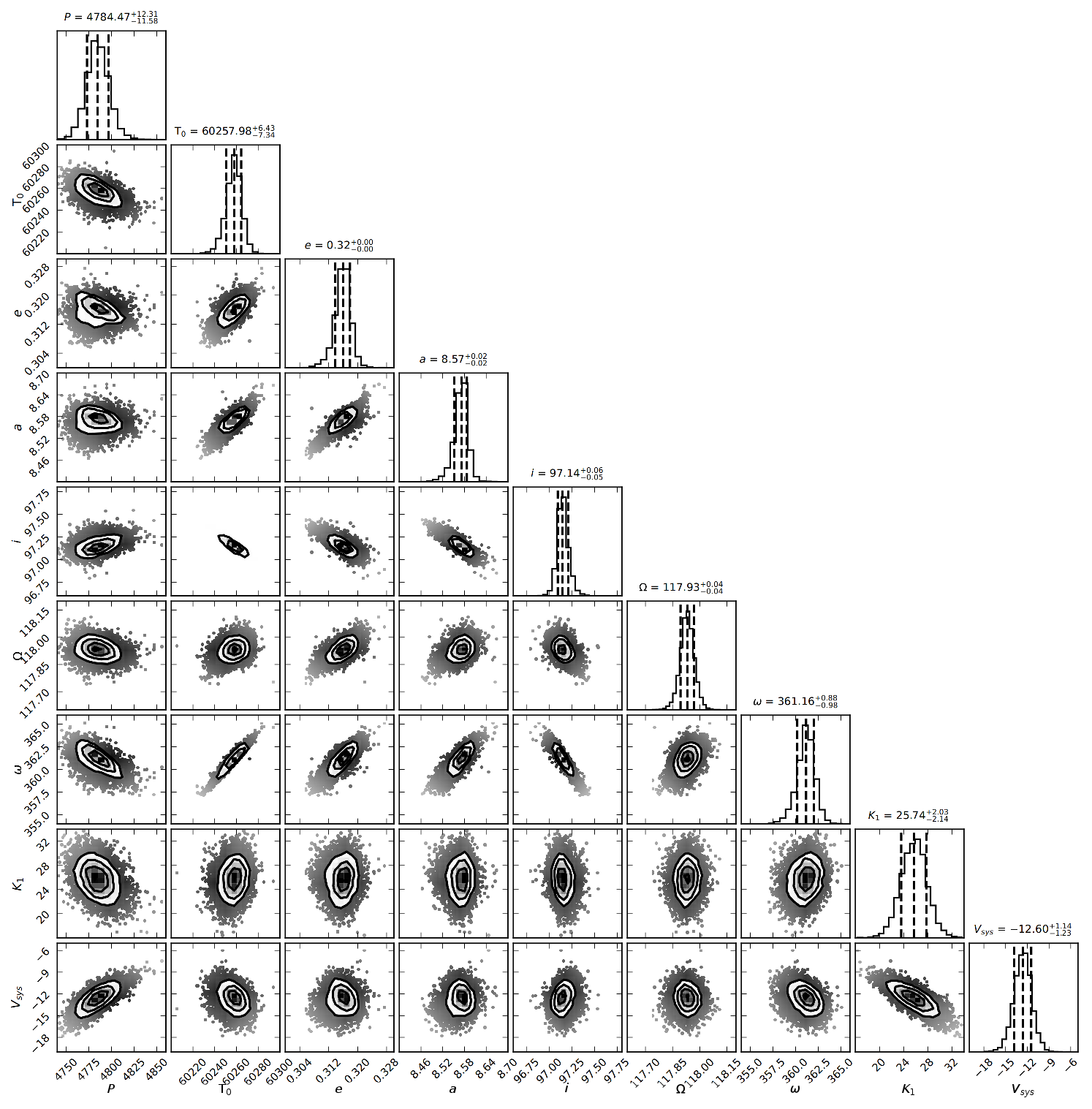}
     \caption{A corner plot showing the interdependencies of fitted parameters in the combined visual and spectroscopic orbital fit. }
     \label{fig:corner}
 \end{figure}

 \figsetstart
\figsetnum{5}
\figsettitle{Binary Fits}

\figsetgrpstart
\figsetend

\begin{figure}
\figurenum{5.9}
  \begin{center}
	\scalebox{0.42}{\includegraphics{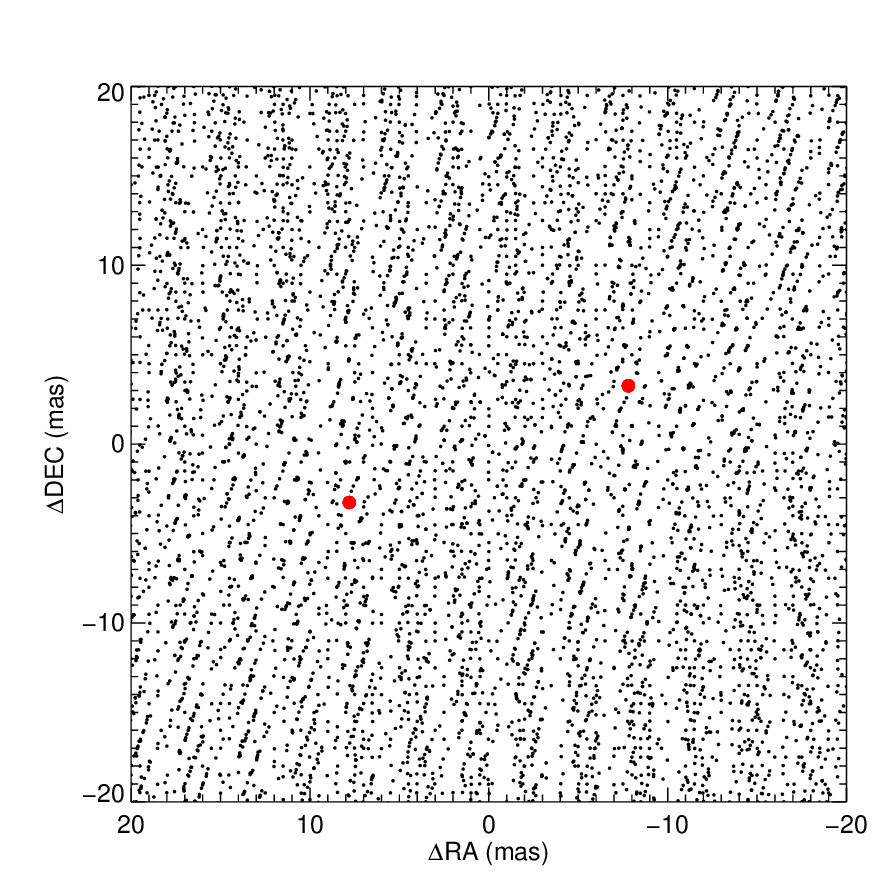}}
	\scalebox{0.54}{\includegraphics{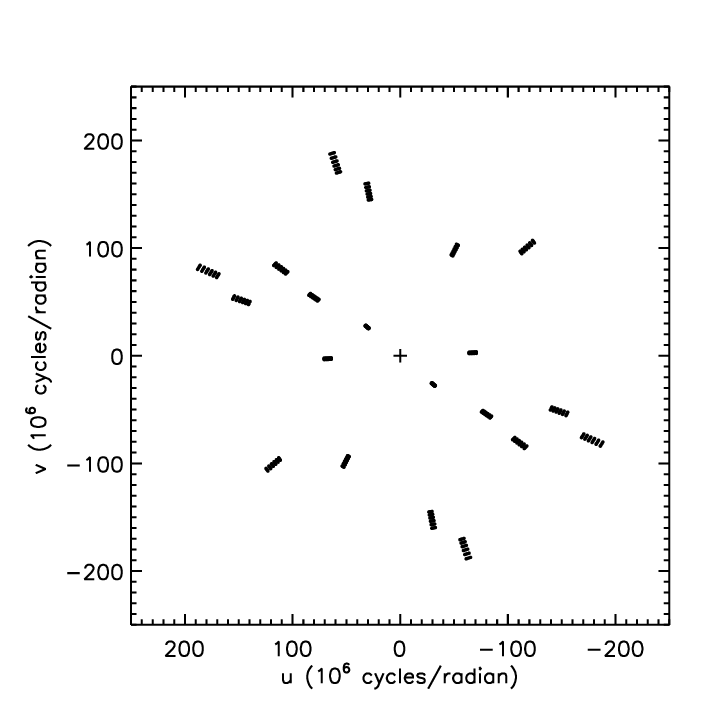}} \\
	\scalebox{0.33}{\includegraphics{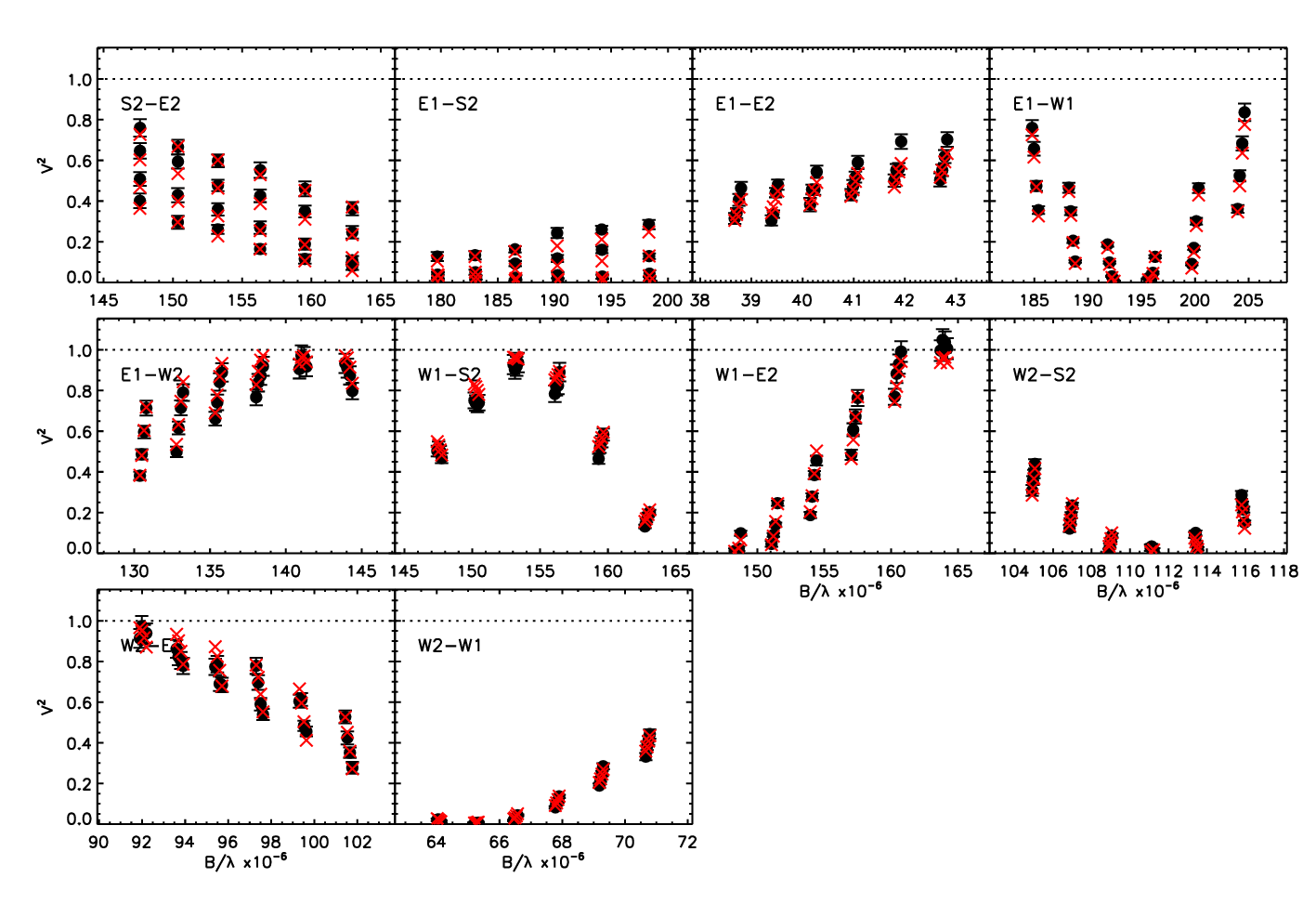}}
	\scalebox{0.33}{\includegraphics{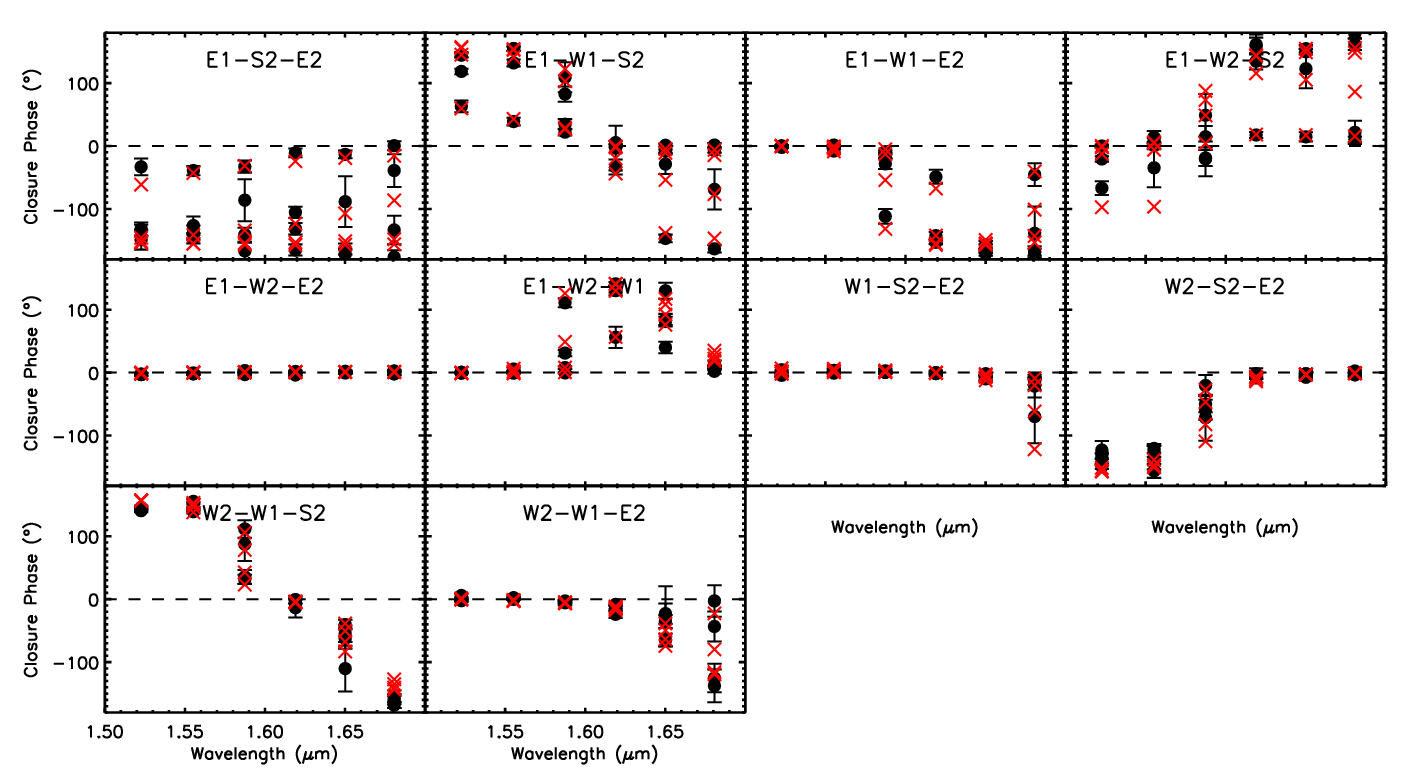}}
        \caption{Top row: $\chi^2$ map (left) from binary fit for WR 137 and $uv$ coverage (right) for MIRC-X data obtained on UT 2019Jul01 (set 2). Bottom row: Visibilities (right) and closure phases (left). Black circles - measured values.  Red crosses - binary fit.}
  \end{center}
\end{figure}

\bibliography{sample631}{}
\bibliographystyle{aasjournal}



\end{document}